\documentclass[12pt, preprint]{aastex}
\begin{document}
\def\simeq{
\mathrel{\raise.3ex\hbox{$\sim$}\mkern-14mu\lower0.4ex\hbox{$-$}}
}

%%%%%%%%%%%%%%%%%%%%%%
%\input epsf.sty
%%%%%%%%%%%%%%%%%%%%%%

%% our macros
\def\lsun{{\rm L_{\odot}}}
\def\msun{{\rm M_{\odot}}}
\def\rsun{{\rm R_{\odot}}} 
\def\lta{\la}
\def\gta{\ga}
\def\be{\begin{equation}}
\def\ee{\end{equation}}
\def\lsun{{\rm L_{\odot}}}
\def\le{{L_{\rm Edd}}}
\def\msun{{\rm M_{\odot}}}
\def\rsun{{\rm R_{\odot}}}
\def\rp{{R_{\rm ph}}}
\def\rs{{R_{\rm s}}}
\def\mo{{\dot M_{\rm out}}}
\def\me{{\dot M_{\rm Edd}}}
\def\tc{{t_{\rm C}}}
\def\rc{{R_{\rm core}}}
\def\mc{{M_{\rm core}}}
\def\mbh{{M_{\rm BH}}}
\def\e{{\dot m_{\rm E}}}

\title{The Brightest Cluster X--ray Sources}

\author{ Andrew~King\altaffilmark{1}}

\altaffiltext{1} {Theoretical Astrophysics Group, University of
Leicester, Leicester LE1 7RH, U.K.; ark@astro.le.ac.uk}

\begin{abstract}

There have been several recent claims of black hole binaries in
globular clusters. I show that these candidate systems could instead
be ultracompact X--ray binaries (UCXBs) in which a neutron star
accretes from a white dwarf. They would represent a slightly earlier
evolutionary stage of known globular cluster UCXBs such as 4U
1820--30, with white dwarf masses $\sim 0.2\msun$, and orbital periods
below 5 minutes. Accretion is slightly super--Eddington, and makes
these systems ultraluminous sources (ULXs) with rather mild beaming
factors $b \sim 0.3$. Their theoretical luminosity function flattens
slightly just above $\le$ and then steepens at $\sim 3\le$. It
predicts of order 2 detections in elliptical galaxies such as NGC
4472, as observed.

The very bright X--ray source HLX--1 lies off the plane of its host
S0a galaxy. If this is an indication of globular cluster membership,
it could conceivably be a more extreme example of a UCXB  with white dwarf
mass $M_2 \simeq 0.34\msun$. The beaming here is tighter
($b \sim 2.5 - 9 \times 10^{-3}$), but the system's distance of
95 Mpc easily eliminates any need to invoke improbable alignment of the
beam for detection. If its position instead indicates membership of a 
satellite dwarf galaxy, HLX--1 could have a much higher accretor mass
$\sim 1000\msun$

\end{abstract}


\section{Introduction}

Globular clusters (GCs) host 10 -- 20\% of the X--ray sources in
galaxies like our own. This fraction rises to $\la 50$\% in
early--type galaxies (e.g. Sarazin et al., 2003). In all galaxies this
implies a far higher incidence of X--ray sources per unit mass in GCs
than in the field -- in the Milky Way by about a factor 100 --
1000. The reason for this is of course that GCs can form close
binaries by dynamical capture, and many of these evolve into low--mass
X--ray binaries (LMXBs). In early--type galaxies this process
continues long after star formation in the field has declined, making
the GC sources more prominent.

A sign of the unusual formation channels available in GCs is the
prominence there of ultracompact X--ray binaries (UCXBs) -- systems
with very short orbital periods $P$ down to $\sim 10$~minutes. For
example, 5 out of the 7 LMXBs with $P < 30$~minutes in the 2011 update
to the catalogue of Ritter \& Kolb (2003) are known to be members of
globular clusters. These systems all appear to involve a neutron star
accreting from a white dwarf companion star. In contrast, systems
containing black holes are rare or absent from GCs. Clearly one would
not expect high--mass X--ray binaries similar to Cygnus X--1, but
there is so far no clear detection in a GC of the soft X--ray
transients which constitute the majority of Galactic black--hole
systems. It is often suggested (e.g. Kalogera, King \& Rasio, 2004)
that most black holes are dynamically expelled from GCs because the
mass contrast with the other stellar populations makes them vulnerable
to the Spitzer mass--segregation instability (Spitzer, 1969).
 
There have nevertheless been a number of claimed or suggested
detections of black--hole X--ray binaries in GCs (Maccarone et al.,
2007; Brassington et al., 2010; Shih et al., 2010; Maccarone et al.,
2011) on the basis of luminosities up to $4.5\times 10^{39}$~erg\,
s$^{-1}$, exceeding the Eddington limit $\le \simeq 10^{38}$~erg\,
s$^{-1}$ for a neutron star, together with $\ga 2\times $ variability
(to rule out superpositions of fainter sources). The same criteria
applied to the much brighter ($\sim 10^{41} - 10^{42}$~erg\, s$^{-1}$)
source HLX--1 have led to the suggestion of a black hole mass $\ga
500\msun$ (Farrell et al., 2009; Wiersema, 2010). HLX--1 is not known
to be a member of a GC: it is associated with an edge--on S0a spiral
galaxy ESO 243--39, lying off the galaxy plane in the outskirts of its
bulge, and so could be a member of a GC or other type of star cluster,
or of a satellite dwarf galaxy.

In the absence of clean dynamical masses ($\ga 3\msun$ for the
claimed GC black holes, and $\ga 500\msun$ for HLX--1) the nature of
these sources is not definitively settled. It is reasonable to ask if
other interpretations are possible, and I attempt this here.

\section{Bright X--ray Sources in Globular Clusters}

The defining feature of the GC black hole candidates is their
luminosity, which is super--Eddington for a neutron star. Any model of
them must therefore allow for high mass transfer rates. Since all the
potential donor stars in GC binaries have low masses $\la 1\msun$ this
obviously also explains the short lifetimes of the black hole
candidates and thus their rarity. We have already noted that the types
of black hole X--ray binaries found in the field -- high--mass
systems, and soft X--ray transients, are not found in GCs. This leaves
one fairly obvious alternative, the ultracompact systems mentioned
above. These systems evolve in a very simple way, which naturally
implies a very high mass transfer rate at a certain epoch (see
e.g. King, 1988 for a discussion). The condition that a white
dwarf donor should fill its Roche lobe leads to
a relation
\begin{equation}
P \sim 0.9m_2^{-1}~{\rm minutes}
\label{mp}
\end{equation}
between donor mass $M_2 = m_2\msun$ and orbital period $P$ (with the
white dwarf taken as a hydrogen--depleted $n = 3/2$ polytrope with
radius--mass relation $R_2 \propto M_2^{-1/3}$).  These very short
orbital periods imply that mass transfer in these UCXB systems is
driven by gravitational radiation, which gives a mass transfer rate
\begin{equation}
-\dot M_2 \simeq 1\times 10^{-3}m_1^{2/3}m_2^{14/3}~\msun\,{\rm
  yr}^{-1}
\label{mdot}
\end{equation}
where $M_1 = m_1\msun$ is the accretor mass. The brightest low--mass
X--ray binary in a Galactic globular cluster is 4U1820--30 in NGC
6624, with $L \simeq (4 - 7)\times 10^{37}$~erg\,s$^{-1}$. This has an
orbital period $P = 11.4$~minutes (Stella, Priedhorsky \& White, 1987;
van der Klis et al., 1993), and so from (\ref{mp}), presumably a
current donor mass $M_2 \simeq 0.08\msun$. Equation (\ref{mdot}) then
gives $-\dot M_2 \simeq 7\times 10^{-9}~\msun\,{\rm yr}^{-1}$, in good
agreement with the observed luminosity if the accretor is a neutron
star of $\simeq 1.4\msun$. The observed X--ray bursts from this system
confirm this identification, and suggest that the donor is a He white
dwarf (Bildsten, 1995; Strohmayer \& Brown, 2002; Cumming, 2003). I
note that Bildsten \& Deloye (2004) consider the effects of chemical
composition and finite entropy and derive slightly more complex
relations than (\ref{mp}, \ref{mdot}), in particular an exponent
closer to 4.13 for $m_2$ in (\ref{mdot}). However their Fig.1 shows
that the deviations are small enough that the poloytropic
approximation is adequate for the purposes of this paper, particularly
for larger values of the white dwarf mass than considered by Bildsten
\& Deloye.

The important result here is that the mass transfer rate could have
been much higher in the past, when $M_2$ was larger. The initial white
dwarf mass in a UCXB is set by its prior evolution. UCXBs presumably
result from some kind of common envelope evolution following the
dynamical capture by the neutron star primary of an evolved companion
star. The degenerate core mass of the companion can range from $\simeq
0.1 - 0.4\msun$ for helium, and up to $\simeq 0.6\msun$ for
carbon/oxygen, depending sensitively on how far the companion has
evolved at the epoch of capture. It is clear that mass transfer rates
well above the Eddington value are easily possible in UCXBs. Note that
$\le$ is twice the usual value since we expect hydrogen--poor
accretion here, i.e.
\begin{equation}
\le \simeq 3.5\times 10^{38}(m_1/1.4))~{\rm
  erg\, s^{-1}}.
\label{leddhe}
\end{equation}
This implies an Eddington accretion rate
\begin{equation}
\me \simeq 5\times 10^{-8}~\msun\, {\rm yr}^{-1}.
\label{mdotedd}
\end{equation}

\section{Ultraluminous X--ray Sources in Globular Clusters}

Compact objects accreting above their Eddington rates probably appear
as ultraluminous X--ray sources (ULXs), and it is now generally
accepted that a large fraction of ULXs are of this type. Disc
accretion in this regime was first described by Shakura \& Sunyaev
(1973). In their picture, radiation pressure becomes important at the
spherization radius $R_{\rm sph} \simeq 27\dot m R_s/4$, where $\dot
m$ is the local accretion rate in units of the Eddington value, and
$R_s = 2GM_1/c^2$ is the Schwarzschild radius of the accretor. Shakura
\& Sunyaev explicitly considered only black hole accretors, but 
their picture also applies to other accretors provided that
$\dot m$ is sufficiently large to ensure $R_{\rm sph} >$~accretor
radius (this always holds for neutron stars with $\dot m >1$ for
example). Inside $R_{\rm sph}$ the disc remains close to the local
radiation pressure limit and blows gas away so that the accretion rate
decreases with disc radius as $\dot M(R) \simeq \dot M(R/R_{\rm sph})
\simeq \me(R/R_s)$. The disc wind has the local escape velocity
at each radius, so mass conservation shows that the wind is dense near
$R_{\rm sph}$ and tenuous near the inner disc edge. The centrifugal
barrier along the disc axis creates a vacuum funnel through which
the luminosity escapes.

Thus super--Eddington accretion produces large apparent X--ray
luminosities through two effects of super--Eddington accretion
(Begelman et al., 2006; Poutanen et al., 2007). First, the bolometric
luminosity is larger than the usual Eddington limit by a factor $\sim
(1 + \ln\dot m)$. Second, the luminosity of a ULX is
collimated by a beaming factor $b$ via scattering off the walls of the
central funnel, so that the apparent luminosity (inferred by assuming
isotropy) is 
\begin{equation}
L_{\rm sph} \simeq {\le\over b}(1 + \ln\dot m).
\label{l}
\end{equation}

Eddington ratios $\dot m >>1$ producing ULXs can arise in a number of
ways in close binary accretion. Thermal--timescale mass transfer from
a massive radiative donor to a lower--mass compact object can give
$\dot m \sim 5000$ or more, as in SS433 (King, Taam \& Begelman, 2000;
Begelman, King \& Pringle, 2006), and nuclear expansion can give
similar rates (Rappaport, Podsiadlowski \& Pfahl,
2005). Super--Eddington rates can arise during disc instabilities
(King, 2002), and I suggest here that UCXBs offer another steady
channel. By contrast, a similar Eddington ratio $\dot m >>1$ is very
difficult to produce in accretion in active galactic nuclei. If the
AGN sits in a spheroid of velocity dispersion $\sigma$ and gas
fraction $f_g$, even the extreme assumption of a full dynamical
accretion rate $\sim f_g\sigma^3/G$, given by suddenly removing
centrifugal support from orbiting gas, produces only $\dot m \la 40$
(King, 2010, eqn 6). Binaries do better than AGN because the
self--gravity of the donor star allows a large gas mass to get close
to the accretor simultaneously by spiralling in through near--circular
orbits. (Dynamical disruption of stars in AGN does not achieve this,
as the likely parabolic orbit means that the arrival time of the gas
is spread out by factors $(M_{\rm SMBH}/M_{\rm star})^{1/2} > 10^3 -
10^4$, cf Lodato et al., 2009.) Accordingly there are no AGN analogues
of ULXs.

Recently, King (2009) noted that the observed {\it inverse}
correlation between soft X--ray emission and blackbody temperature
($L_{\rm soft} \propto T_{bb}^{-3.5}$) for some ULXs (Kajava \&
Poutanen, 2009) could be understood if the beaming factor varies as
\begin{equation}
b \sim {73\over \dot m^2},
\label{beam}
\end{equation} 
a form which also follows from simple geometrical arguments about the
funnel opening angle. (We note that this suggests that significant
beaming occurs only for sufficiently super--Eddington accretors $\dot
m \ga 8.5$.)  Mainieri et al (2010) used this form of beaming to study
the luminosity function of a sample of ULXs out to redshifts $z = 0.3$, finding
excellent agreement. 

Accordingly, using (\ref{leddhe}, \ref{l}) I adopt 
\begin{equation}
L_{\rm sph} = 4.4\times 10^{36}m_1\dot m^2(1 + \ln \dot m)~{\rm
  erg\, s^{-1}}.
\label{lum}
\end{equation}
for the apparent UCXB luminosity for $\dot m > 8.5$, setting $b=1$ in
(\ref{l}) for $1 < \dot m < 8.5$. 

A super--Eddington UCXB system with a $1.4\msun$ neutron star
accretor and an Eddington factor $\dot m \simeq 15$ has a $L_{\rm sph}
\simeq 5 \times 10^{39}~{\rm erg\, s^{-1}}$, large enough to explain
all the putative black--hole systems in GCs. The beaming factor here
is $b \simeq 0.32$. To see if this is reasonable we need to work out
the luminosity function of these systems.

\section{Luminosity Function}

Bildsten \& Deloye (2004) note that the evolution of UCXBs is so rapid
that one can derive their GC luminosity function (LF) by assuming staeady
state conditions. This gives
\begin{equation}
{{\rm d}N\over {\rm d}L} \propto L^{-\alpha}.
\label{lf}
\end{equation}

Extending the LF to luminosities where beaming is important is
straightforward, following Bildsten \& Deloye (2004). They note that
UCXBs evolve through the observable range of luminosities on a
timescale far shorter than the age of GCs or any reasonable estimate
of the time for their birthrate to change. Hence if we know the time
$t$ that a UCXB spends above a given luminosity, then the
cumulative LF $N(>L)$ is directly proportional to $t$. 

To find $t$ as a function of $L$ I integrate the evolution equation
(\ref{mdot}) assuming $M_2 << M_1$, as $t \propto M_2^{1 - \beta}$,
where I have generalized the exponent of $m_2$ in (\ref{mdot}) from
the polytropic value 14/3 to a general $\beta$ to enable comparison
with Bildsten \& Deloye, and also assumed that the current donor mass
is much smaller than its initial value. This gives $t \propto (-\dot
M_2)^{-(\beta - 1)/\beta}$. If there were no beaming, we would now
replace $-\dot M_2$ by $L$ and get $N(>L) \propto t \propto L^{-(\beta
  - 1)/\beta}$ and so ${\rm d}N/{\rm d}L \propto L^{1/\beta - 2}$ for
$\dot m < 1$ (Bildsten \& Deloye, 2001).

However beaming changes both the connection
between $-\dot M_2$ and $L$, and also the number of systems we can
see: we now have $L \propto b^{-1} \propto \dot m^2$ (neglecting the
logarithmic dependence on $\dot m$, and $N(>L) \propto t\cdot b
\propto \dot m^{-(\beta - 1)/\beta -2}$. Together these give $N(>L)
\propto L^{-(\beta - 1)/2\beta -1}$, and
\begin{equation}
{{\rm d}N\over {\rm d}L} \propto L^{1/2\beta  -5/2}
\label{lf2}
\end{equation}
for $\dot m > 8.5$.
If indeed there is essentially no beaming for $1 < \dot m < 8.5$ as suggested 
just below eqn (\ref{lum}), the logarithmic relation between 
$L$ and $\dot m$ implies that are more UCXBs than naively expected 
between $L = \le$ and $L \simeq 3\le$, so the luminosity function
flattens there, i.e.
\begin{equation}
{{\rm d}N\over {\rm d}L} \propto \exp 0.77\biggl(1 - {L\over \le}\biggr).
\label{lf3}
\end{equation}
Enforcing continuity at the breaks $L = \le, 3.14\le$, the UCXB
luminosity function is 
\begin{equation}
l^{-1.77}, \ \ \ \exp[0.77(1-l)], \ \ \ 2.93
l^{-2.38}
\end{equation}
over these ranges, where we have taken $\beta = 4.3$ and set $l = L/\le$.

\section{Candidate GC Black Holes}

Maccarone et al. (2007, 2011) discuss two black hole candidate sources
in the elliptical galaxy NGC 4472. They are XMMU 122939.7+075333 ($L
\la 4.5\times 10^{39}$~erg\, s$^{-1}$) and CXOU 1229410+075744 ($L \la
2\times 10^{39}$~erg\, s$^{-1}$). The LF of Kim \& Fabbiano (2004) for
NGC 4472 shows that there are some $\sim 300$ X--ray sources with
$4\times 10^{37}~{\rm erg\,s^{-1}} < L < L_B$ (their Fig.2: note that
the displayed LF is scaled upwards by a factor 29.51 for display
purposes). If a substantial fraction of these are UCXBs, as expected,
the theoretical LF (\ref{lf2}, \ref{lf3}) shows that the presence of
two ultracompact {\it neutron--star} ULXs with the quoted luminosities
is perfectly reasonable. For the more extreme (XMMU) source the white
dwarf companion has mass $\simeq 0.2\msun$, the orbital period is $P
\simeq 4.5$~minutes, and the beaming factor is $b \simeq 0.32$.

\section{HLX--1}

Soria et al. (2010) find a red optical counterpart to HLX-1. At the
distance to ESO 243-49 this has a luminosity compatible with a massive
globular cluster. On the other hand this luminosity is also compatible
with the nucleus of a stripped dwarf galaxy, and in addition Wiersema
et al., (2010) find a bright H$\alpha$ line associated with the
source. If this comes directly from the accretion flow it rules out a
UCXB model, although one has still to rule out an origin in a nebula
around the source -- this appears unlikely in a gas--poor environment
like a globular cluster. In view of these uncertainties I consider a
number of possibilities.

If HLX--1 is a UCXB in a globular cluster we can use
(\ref{lum}) with $m_1 = 1.4$ to find an estimated Eddington ratio
$\dot m \simeq 170$, with a beaming factor $b \simeq 2.5 \times
10^{-3}$, which would require a white dwarf donor mass $M_2 =
0.34\msun$, and an orbital period $P \simeq 2.5$~minutes. As a check
we can estimate the distance $D_{\rm min}$ to the nearest object
plausibly observable with this amount of beaming (i.e. without
requiring an improbable `aiming' of the beam at Earth).  This is given
by eqn (12) or (15) of King (2009) (but note that the coefficient in
the final form of eqn. (15) should be 258 and not 660).  This gives
$D_{\rm min} \simeq 17N^{-1/3}$~Mpc, where $N$ (presumably $\sim 1$)
is the number of objects of this type in the host galaxy. This is
comfortably smaller than the known distance of 95~Mpc, so it is not
unreasonable to be in the beam of this system.

If instead of being a globular cluster UCXB, HLX-1 involves $10\msun$
black hole accreting hydrogen--rich material we halve the coefficient
on the rhs of (\ref{lum}) and set $m_1 =10$. This gives an Eddington
ratio $\dot m=90$, which is quite possible for either a high--mass
X--ray binary or a soft X--ray transient. The beaming factor is $b
\simeq 9\times 10^{-3}$, and the minimum reasonable distance is
$D_{\rm min} \simeq 11N^{-1/3}$~Mpc. Finally, if HLX--1 is actually in
a dwarf galaxy interacting with ESO 243--39, it is quite possible that
the accretor is the central massive black hole of this galaxy with a
mass $\ga 1000\msun$. In this case the interaction with the larger
galaxy can trigger accretion (King \& Dehnen, 2005). Lasota et
al. (2011) consider other possibilities of this type.

\section{Conclusions}

This paper has shown that the claimed black hole binaries in globular
clusters are also understandable as ultracompact X--ray binaries in
which a neutron star accretes from a white dwarf. The high luminosity
arises from the higher mass transfer rates expected from progenitors
of known UCXBs such as 4U 1820--30, with white dwarf masses $\sim
0.2\msun$. The theoretical luminosity function of these systems agrees
with the detected number. UCXBs have orbital periods of only a few
minutes. Detection of such periods would strongly support a UCXB
model, while measurement of secure dynamical masses would break the
ambiguity between neutron star and black hole accretors. However if
the systems are indeed UCXBs both of these will prove difficult, as
the mild beaming induced by super--Eddington accretion suggests that
the systems are likely to be face--on. Longer superorbital periods may
be seen as a result of precessions in either case. Similarly, it will
be difficult to decide spectroscopically whether the accreting
material is hydrogen--depleted or not, as one needs to find a line
feature securely associated with the binary.

The very bright X--ray source HLX--1 could conceivably be a more
extreme example of a UCXB (white dwarf mass $M_2 \simeq 0.34\msun$) if it
is indeed a member of a globular cluster. However if it is a member of a 
dwarf satellite galaxy its mass could be $\sim 1000\msun$. There is currently
no easy way to eliminate any of these possibilities.

\acknowledgments

Research in theoretical astrophysics at Leicester
is supported by an STFC Rolling Grant.

\end{document}